\documentclass[12pt]{article}
\usepackage[english]{babel}
\usepackage{psfig,epsfig,bezier}
\begin{document}
\def\bsig{\mbox{\boldmath $\sigma$}}                          
\def\bsig{\mbox{\boldmath $\Sigma$}}
\def\bgam{\mbox{\boldmath $\gamma$}}
\def\bgam{\mbox{\boldmath $\Gamma$}}
\def\bphi{\mbox{\boldmath $\phi$}}
\def\bphi{\mbox{\boldmath $\Phi$}}
\def\btau{\mbox{\boldmath $\tau$}}
\def\btau{\mbox{\boldmath $\Tau$}}
\def\btau{\mbox{\boldmath $\partial$}}
\def\Delc{{\Delta}_{\circ}}
\def\bp{\mid {\bf p} \mid}
\def\al{\alpha}
\def\bet{\beta}
\def\gam{\gamma}
\def\del{\delta}
\def\Del{\Delta}
\def\te{\theta}
\def\nua{{\nu}_{\alpha}}
\def\nui{{\nu}_i}
\def\nuj{{\nu}_j}
\def\nue{{\nu}_e}
\def\num{{\nu}_{\mu}}
\def\nut{{\nu}_{\tau}}
\def\2te{2{\theta}}
\def\chic#1{{\scriptscriptstyle #1}}
\def\chicl{{\chic L}}
\def\lam{\lambda}
\def\SU{SU(2)_{\chic L} \otimes U(1)_{\chic Y}}
\def\Lam{\Lambda}
\def\sig{\sigma}
\def\'#1{\ifx#1i\accent19\i\else\accent19#1\fi}
\def\O{\Omega}
\def\o{\omega}
\def\s{\sigma}
\def\D{\Delta}
\def\d{\delta}
\def\df{\rm d}
\def\8{\infty}
\def\ld{\lambda}
\def\eps{\epsilon}
\def\theequation{\thesection.\arabic{equation}}
\newcommand{\be}{\begin{equation}}
\newcommand{\ee}{\end{equation}}
\newcommand{\ba}{\begin{array}}
\newcommand{\ea}{\end{array}}
\newcommand{\dis}{\displaystyle}
\newcommand{\alfad}{\frac{\dis \bar \alpha_s}{\dis \pi}}
\newcommand{\bra}{\mbox{$<$}}
\newcommand{\ket}{\mbox{$>$}}
\def\Frac#1#2{\frac{\displaystyle{#1}}{\displaystyle{#2}}}
\thispagestyle{empty}
\begin{titlepage}
\begin{center}
\vskip 1.5cm
{\LARGE \bf Hydrodynamics of Galactic Dark Matter}
\end{center}
\normalsize
\vskip1cm
\begin{center}
{\large \bf Luis G. Cabral-Rosetti}$^a$
\footnote{E-mail: luis@nuclecu.unam.mx},
{\large \bf Tonatiuh Matos}$^b$
\footnote{E-mail: Tonatiu.Matos@fis.cinvestav.mx}, \\
{\large \bf Dar{\'\i}o Nu\~nez}$^{a,c}$
\footnote{E-mail: nunez@nuclecu.unam.mx and nunez@gravity.phys.psu.edu}
{\large and} {\large \bf Roberto A. Sussman}$^{a}$
\footnote{E-mail: sussman@nuclecu.unam.mx and sussky@edsa.net.mx}
\end{center}
\begin{center}
\baselineskip=13pt
$^a${\large \it Instituto de Ciencias Nucleares,\\
Universidad Nacional Aut\'onoma de M\'exico, (ICN-UNAM).}\\
\baselineskip=12pt
{\it Circuito Exterior, C.U., Apartado Postal 70-543, 94510 M\'exico,
D.F.,  M\'exico.}\\
\vskip0.5cm
$^b${\large \it Departamento de F{\'\i}sica,\\
Centro de Investigaci\'on y de Estudios Avanzadas del IPN,}\\
\baselineskip=12pt
{\it Apartado Postal 14-740, M\'exico D. F., M\'exico.}\\
\vskip0.5cm
$^c${\large \it Center for Gravitational Physics and Geometry, \\
Penn State University,}\\
\baselineskip=12pt
{\it University Park, PA 16802, U.S.A.}\\
\vglue 0.8cm
\end{center}


\date{\today}
%
\begin{abstract}
We consider simple hydrodynamical models of galactic dark matter in which 
the galactic halo is a self-gravitating and self-interacting gas that 
dominates the dynamics of the galaxy. Modeling this halo as a sphericaly 
symmetric and static perfect fluid satisfying the field equations of General 
Relativity, visible barionic matter can be treated as  ``test particles'' 
in the geometry of this field.  We show that the assumption of an empirical 
``universal rotation curve'' that fits a wide variety of galaxies is 
compatible, under suitable approximations, with state variables 
characteristic of a non-relativistic Maxwell-Boltzmann gas that becomes 
an  isothermal sphere in the Newtonian limit. Consistency criteria lead
to a minimal bound for particle masses in the range 
$30 \,\hbox{eV} \leq m \leq 60 \,\hbox{eV}$ and to a constraint between the 
central temperature and the particles mass. The allowed mass range includes 
popular supersymmetric particle candidates, such as the neutralino, axino and 
gravitino, as well as lighter particles ($m\approx$ keV) proposed by numerical
N-body simulations associated with self-interactive CDM and WDM  structure 
formation theories.  
\end{abstract}

\noindent{\em PACS:}
      95.35.+d, 
      95.30.Sf, 
      98.80.-k  
\vfill

\end{titlepage}

\setcounter{equation}{0}
\section{Introduction}

\noindent
The presence of large amounts of dark matter at the galactic lengthscale is 
already an established fact. Assuming that this dark matter is a gas (or gas 
mixture) of various particles species, the established classification 
criteria labels possible dark matter forms as ``cold'' or ``hot'', depending 
on the relativistic or non-relativistic nature of the particles' energetic 
spectrum at their decoupling from the cosmic mixture \cite{Trimble}, 
\cite{tbook_1}, \cite{tbook_2}. Hot dark matter (HDM) scenarios seem to be 
incompatible with current theories of structure formation and thus, are not 
favoured dark matter candidates \cite{tbook_1}, \cite{tbook_2}, 
\cite{dm_scenarios}. Cold dark matter (CDM), usualy examined within a 
newtonian framework, can be considered as non-interactive (a self gravitating 
gas of collisionless particles) or self-interactive \cite{old_cdm_theo}. CDM 
models are often developed in terms of n-body numerical simmulations 
\cite{old_cdm_nbody}, \cite{nbody_1}, \cite{nbody_2}, \cite{nbody_3}. 
Non-interactive CDM models present the following discrepancies with 
observations at the galactic scale \cite{cdm_problems_1}, 
\cite{cdm_problems_2}: (a) the ``substructure problem'' related to excess 
clustering on sub-galactic scales, (b) the ``cusp problem'' characterized by 
a monotonic increase of density towards the center of halos, leading to 
excessively concentrated cores. These problems appear in the more recent 
numerical simulations (see \cite{nbody_1}, \cite{nbody_2}, \cite{nbody_3}). 
In order to deal with these problems, the possibility of self-interacting 
dark matter has been considered, so that nonzero pressure or thermal effects 
can emerge, thus leading to self-interactive models of CDM ({\it i.e.} 
SCDM) \cite{scdm_1}, \cite{scdm_2}, \cite{scdm_3}, \cite{scdm_4}, 
\cite{scdm_5} and ``warm'' dark matter (WDM) models \cite{wdm_1}-\cite{wdm_6} 
that challenges the duality CDM vs. HDM. Other proposed dark matter sources 
consist replacing the gas of particles approach by scalar fields \cite{sfe_1},
\cite{sfe_2} and even more ``exotic'' sources \cite{sfe_3}. 

Whether based on SCDM or WDM, current theories of structure formation point 
towards dark matter characterized by particles having a mass of the order of 
at least keV's (see \cite{scdm_1}-\cite{wdm_6}), thus suggesting that massive 
but light particles, such as electron neutrinos and axions (see Table~1), 
should be eliminated as primary dark matter candidates (though there is no 
reason to assume that these particles would be absent in galactic halos). Of 
all possible particle candidates (denoted as WIMP's: weakly interactive 
massive particles) complying with the required mass value of relique gases, 
only the massive neutrinos (the muon or tau neutrinos), have been detected, 
whereas other WIMPS (gravitino, sterile neutrino, axino, etc.) are 
speculative. See \cite{Pal}, \cite{SNO}, \cite{PDG}, \cite{Ellis} and Table~1 
for a list of candidate particles and appropriate references.

Even if the dynamics of visible matter in galaxies can be described
succesfuly with Newtonian gravity, we believe that General Relativity 
is an appropriate framework for understanding basic features of 
galactic dark matter, a gravitational field source whose precise physical 
nature still remains  an open question.  If the results obtained with 
GR coincide with Newtonian results, then there is no harm done from 
a pragmatic calculations-oriented point of view. However, 
from a formal-theoretical approach, we believe it is beneficial 
to broaden the scope of the study of galactic dynamics by incorporating 
it to a more general gravitational theory.  In particular, in this paper 
we aim at testing the compatibility between observed galactic rotation 
curves and simple thermodynamical assumptions under the framework of GR.

Since dark matter halo probably constitutes the overwhelming mayority (90 \%) 
of the galactic mass, an alternative approach to numerical simulations and 
newtonian hydrodynamics follows by a general relativistic model describing 
the gravitational field of the galaxy as a spacetime geometry generated by 
the the dark matter halo (as a self gravitating gas), hence visible matter 
becomes test particles that evolve along stable geodesics of this spacetime. 
There is strong empiric evidence that the radial profile of rotational 
velocities (``rotation curves'') in most galaxies roughly fits a ``universal 
rotation curve'' (URC) \cite{urc_1}, \cite{urc_2}. This URC is characterized 
by a ``flattening'' effect whereby rotation velocities tend to a constant 
``terminal'' velocity whose value depends on the type of galaxy (between 
$125\,\hbox{km/sec}$ and $250\,\hbox{km/sec}$). The profile of rotation curves 
identifies two main contributions of galactic matter: visible matter (the 
disk), showing a keplerian decay, and dark matter (the halo), explaining the 
flattening effect. This kinematic evidence might allow us to determine
(at  least partialy) the geometry of the spacetime associated with a self 
gravitating galaxy. In other words,  our approach somehow  inverts the
standard initial value procedure in general relativistic  hydrodynamics:
instead of prescribing initial data based on physicaly  motivated sources
and then find the geometry of spacetime and the trayectories of test
particles after solving Einstein's equations, we provide first
constraints on the  geometry of spacetime (from symmetry criteria and
empirical kinematic data)  and then find, with the help of the field
equations, the corresponding  momentum-energy tensor of the sources. This
approach to galactic dark matter  has been used in connection to scalar
fields
\cite{sfe_1}.

Bearing in mind that the dark matter halo overwhelmingly dominates the 
galactic matter content (at least in the halo region), we shall assume that 
the galactic halo (as a self gravitating gas) is the unique matter source of 
the galactic spacetime. Visible matter becomes then test observers that 
follow stable circular geodesic orbits (the galactic rotation curves) of this
spacetime. Following the ``inverse'' approach described above, we propose to 
use the empiric law governing the form of the URC for the galactic halo (see 
\cite{urc_1} and \cite{urc_2}) in order to make specific asertions on the 
nature of the sources of the galactic spacetime. Considering the 
self-gravitating galactic halo gas to be self-interactive (instead of 
colissionless matter or a scalar field), we aim at verifying if the assumption
of the URC profile for the rotation velocity of geodesic observers (rotation 
curves) is compatible with the assumption that the galactic halo gas is a 
simple self-gravitating and self-interactive gas in thermodynamical 
equilibrium. For this purpose, we consider the galactic halo to be a 
spacetime characterized as: (a) sphericaly symmetric, (b) its energy-momentum 
tensor is that of a perfect fluid satisfying the equation of state of an 
equilibrium Maxwell-Boltzmann gas in its non-relativistic limit \cite{RKT}, 
\cite{rund}. Assumption (a) is supported by observations in the halo of 
galaxies, while (b) is the central hypothesis in the present
work.  Assumption (b) requires spacetime to be stationary (static if
rotation vanishes) and leads to a law relating temperature  gradients with
the 4-acceleration (Tolman Law). Then, since we are assuming  the
validity of the empiric URC for the galactic halo, we need to cast the 
field equations and the conditions imposed by the thermodynamics in terms
of  this rotation velocity ({\it i.e.} the velocity of test particles in
stable  geodesic orbits), considered now as a dynamical variable. Using
the URC  empiric law as an {\it ansatz} for this velocity immediately
leads to  expressions for the state variables that are (under suitable
approximations)  consistent with the thermodynamics of the
non-relativistic Maxwell-Boltzmann  ideal gas. From these expressions and
bearing in mind numerical estimates of  the empiric parameters appearing
in the URC (``terminal'' rotation velocities  and the ``core radius''),
we obtain: (1) a constraint on the ratio of the  particles mass to
temperature for this gas, (2) the criterion of applicability of the
Maxwell-Boltzmann distribution ({\it i.e.} the non-degeneracy criterion)
\cite{Landau}, leading to a minimal  bound of about $30$ to $60$ eV for
the mass of the gas particles. Therefore,  the assumption of a
Maxwell-Boltzmann gas (SCDM or WDM) model for the  galactic halo leads
to an acceptable value for the particle's mass lying in  the range
$m > 0.5\,\hbox{keV}$. We provide in Table~1 a list  of particle candidates 
that could be accomodated according to the criteria (1) and (2) above, namely:
neutralino, photino, light gravitino, sterile neutrino, dilaton, axino, 
majoron, mirror neutrino and possibly standard massive neutrinos. As 
mentioned previously, this mass range is compatible with predictions of 
current work based on SCDM and WDM structure formation models. We find it 
interesting to remark that barions and electrons comply with the criterion 
(2) above, but (1) would imply gas temperatures of the order of
$10^{3}-10^{6}$ K. A gas of barions or electrons at such temperatures
would certainly not be  ``dark''. HDM or WDM models based on less massive
particles, like the electron  neutrino, remain outside the
scope of the present work, since these particles might require
assuming a fuly relativistic Maxwell-Boltzmann gas or  a degenerate gas
(possibly relativistic) complying with a Fermi-Dirac or  Bose-Einstein
statistics. The axion, as well as other non-thermal relique sources, are
also outside the scope of this paper and their study requires a different
approach.    

The paper is composed as follows. In the next section we present the field 
equations for a static sphericaly symmetric spacetime with a perfect fluid 
source. We provide in section 3 a review of the thermodynamics of an 
equilibrium Maxwell-Boltzmann gas. Then, in section 4, we re-write the 
field equations in terms of the orbital velocity of stable circular geodesics 
and then assume for this velocity the empiric {\it ansatz} given by the URC. 
This leads to forms of the state variables that will be compatible, under 
suitable series approximations, with the thermodynamics of the 
Maxwell-Boltzmann gas. Putting all these results together, we discuss in
the  last section the possible ranges for the mass of the particles of the
dark  halo gas and suggest future lines of research.

\setcounter{equation}{0}
\section{Field Equations}

As mentioned above, we consider the halo to be spherically symmetric and 
is the determining component of the stress energy tensor determining the 
geometry. Thus, we consider the line element of an spherically symmetric 
space time:

\begin{equation}
ds^{2}\ =\ -B^{2}(r)\,c^{2}\,dt^{2}\ +\ \frac{d\,r^{2}}{1-2\,M(r)/r}\ +\ 
r^{2}\,\left( d\theta ^{2}+\sin ^{2}\theta \,d\varphi ^{2}\right) . 
\label{eq:ele0}
\end{equation}

\noindent
Assuming as the source of such line element a static perfect fluid momentum 
energy tensor $T^{ab}=(\rho +p)\,u^{a}u^{b}+p\,g^{ab}$, with 
$u^{a}=B^{-1}\,\delta ^{a}\,_{ct}$, we obtain the following field equations

\begin{equation}
-G^{t}\,_{t}\ =\ \kappa \,\rho \ =\ \frac{2\,M^{\prime }}{r^{2}},
\label{eq:rho1}
\end{equation}

\begin{equation}
G^{r}\,_{r}\ =\ \kappa \,p\ =\ \frac{2}{r}\,\left( 1-\frac{2\,M}{r}\right) \,
\frac{B^{\prime }}{B}\ -\ \frac{2\,M}{r^{3}},  
\label{eq:p1}
\end{equation}

\begin{eqnarray}
&&G^{r}\,_{r}-G^{\theta }\,_{\theta }\ =\ 0\Rightarrow   \nonumber \\
&&\left( 1-\frac{2\,M}{r}\right) \,\frac{B^{\prime \prime }}{B}\ -\ \frac{1}
{r}\,\left( 1+M^{\prime }-\frac{3\,M}{r}\right) \,\frac{B^{\prime }}{B}\ +\
\frac{3\,M}{r^{3}}\ -\ \frac{M^{\prime }}{r^{2}}\ =\ 0,  
\label{eq:P1}
\end{eqnarray}

\noindent
where $\kappa =8\pi G/c^{4}$ and a prime denotes derivative with respect to 
$r$. As mentioned in the Introduction (see also \cite{sfe_1} and
\cite{sfe_2}), it is possible (by working  with Einstein's equations
backwards) to impose constrains on the geometry of  the spacetime that
yield valuable information regarding the type of matter sources curving
such spacetime. We apply this reasoning to the observed velocity  profile
of stars orbiting around a galaxy, considered as test particles  moving in
stable circular geodesics. Knowing the specific form of the  velocity
profiles around these geodesics should allow us to infere at least  basic
features on the nature sources producing the galactic field. For a 
sphericaly symmetric spacetime (\ref{eq:ele0}), the energy, $E$, and the 
angular momentum,
$L$, are conserved quantities for any particle moving in a  geodesic,
hence the geodesic equation for the radial motion has the following  form:

\begin{equation}
\dot{r}^{2}-{\frac{1}{{B^{2}}}}\left( {\frac{{E^{2}}}{{B^{2}}}}
-{\frac{{L^{2}}}{{r^{2}}}}-1\right) = 0 \ ,
\end{equation}

\noindent
where a dot denotes derivative with respect to the affine parameter of the 
geodesic. For a test particle to be in stable circular geodesic motion in 
any static spherically symmetric space time, its energy and angular momentum 
must satisfy:

\begin{eqnarray}
E^{2} &=&{\frac{{B^{3}}}{{B-rB^{\prime }}}}\ , \\
L^{2} &=&{\frac{{r^{3}\,B^{\prime }}}{{B-r\,B^{\prime }}}}\ .
\label{eq:ELes}
\end{eqnarray}

\noindent
The tangential velocity of these test particles, $V(r)$, can also be 
expressed in terms of the metric coefficients as:

\begin{equation}
\frac{V^{2}}{c^{2}}\ \equiv \ v^{2}(r)\ =\ {\frac{{r\,B^{\prime }}}{{B}}}.
\label{eq:vt}
\end{equation}

In previous work (see \cite{sfe_1}, \cite{sfe_2} and \cite{sfe_3}) this 
tangential velocity was assumed to be constant along the full domain of the 
solution. In the present paper we consider $v=v(r)$ and eliminate the metric 
coefficient, $B(r)$, and its derivatives in (\ref{eq:rho1})-(\ref{eq:P1}) in 
terms of the tangential velocity, $v^{2}(r)$, as given by Eq.(\ref{eq:vt}). 
This leads to a form for the field equations in which $v(r)$ becomes a 
dynamical variable replacing $B(r)$.

After some algebraic manipulation, equation (\ref{eq:P1}) becomes the 
following constraint relating the metric function $M$ and the tangential 
velocity:

\begin{equation}
M^{\prime}\ + \ {\frac {\left (-3-5\,v^2+4\,vv^{\prime}\,r+2\,{v}^{4}\right
)M} {r\left (1+{v}^{2}\right )}} \ - \ {\frac {v\left
(-2\,v+2\,v^{\prime}\,r+{v}^{3} \right )}{1+{v}^{2}}} \ = \ 0.
\label{eq:const}
\end{equation}

\noindent
Substitution of this last equation into equations (\ref{eq:rho1}) and 
(\ref {eq:p1}) provides the following expressions for the density and the 
pressure of the fluid in terms of $M$, $v$ and $v^{\prime}$:

\begin{eqnarray}
\kappa \,p &\ =\ &2\,{\frac{-M-2\,M{v}^{2}+{v}^{2}r}{{r}^{3}}}, \\
\kappa \,\rho  &\ =\ &{\frac{\left[ -8\,vv^{\prime }\,r-2\,\left( 2\,
{v}^{2}+1\right) \left( {v}^{2}-3\right) \right] M+4\,{r}^{2}vv^{\prime }
+2\,{v}^{2}\left( -2+{v}^{2}\right) r}{{r}^{3}\left( 1+{v}^{2}\right) }}.  
\nonumber
\\
&&  \label{eq:rp}
\end{eqnarray}

It is remarkable to see how the replacement of $B$ by $v$ considerably 
simplifies the field equations for a general static and sphericaly symmetric 
field with a perfect fluid source. Writing the field equations in terms of 
the orbital velocity, $v$, provides a useful insight into how an (in 
principle) observable quantity relates to spacetime curvature and with 
physical quantities (state variables) which characterize the source of 
spacetime. Thus, given an empirical functional form for $v(r)$ (a rotation
``profile'' for test particles), we can obtain $M(r)$ by integrating the
constraint (\ref{eq:const}) and thus, we arrive to fully determined forms of
$\rho $ and $p$ in (\ref{eq:rp}).

\setcounter{equation}{0}
\section{Thermodynamics}

We aim at verifying if the empiric laws associated with galactic rotation 
curves (hence, associated with $v(r)$) can be compatible with matter sources 
that satisfy basic physical considerations and principles. Since galactic 
dark matter is, most probably, non-relativistic and the assumption of a 
perfect fluid source for the static metric (\ref{eq:ele0}) points to an 
equilibrium configuration, it is tempting to verify if $\rho$ and $p$ 
associated with a given empirical form for $v(r)$ correspond to state 
variables characteristic of simple, non-relativistic systems in 
thermodynamically equilibrium, such as a suitable ideal gas in its 
non-relativistic limit \cite{RKT}, \cite{rund}.

If we assume that the self gravitating ideal ``dark'' gas exists in physical 
conditions far from those in which the quantum properties of the gas 
particles are relevant, we would be demanding that these particles comply 
with Maxwell-Boltzmann (MB) statistics. Following \cite{Landau}, the 
``non-degeneracy'' condition that justifies an MB distribution is given by

\begin{equation}
\frac{n\,\hbar ^{3}}{(m\,k_{_{B}}\,T)^{3/2}}\ll 1\ ,  
\label{eq:landau}
\end{equation}

\noindent
where $n$, $T$, $\hbar $ and $k_{_{B}}$ are, respectively, the particle 
number density, absolute temperature, Planck's and Boltzmann's constants. If 
the constraint (\ref{eq:landau}) holds and we further assume thermodynamical 
equilibrium and non-relativistic conditions, the ideal dark gas must satisfy 
the equation of state of a non-relativistic monatomic ideal gas

\begin{equation}
\rho \ =\ mc^{2}\,n\ +\ \frac{3}{2}\,n\,k_{_{B}}T,\qquad \qquad p\ =\
n\,k_{_{B}}T,  
\label{eq:mb}
\end{equation}

\noindent
whose macroscopic state variables can be obtained from a MB distribution 
function under an equilibrium Kinetic theory approach (the non-relativistic 
and non-degenerate limit of the J\"{u}ttner distribution) \cite{RKT}. An 
equilibrium MB distribution restricts the geometry of spacetime \cite{rund}, 
resulting in the existence of a timelike Killing vector field 
$\beta^{a}=\beta \,u^{a}$, where $\beta \equiv mc^{2}/k_{_{B}}T$, as well as 
the following relation (Tolman's law) between the 4-acceleration and the 
temperature gradient

\begin{equation}
\dot{u}_{a}\ +\ h_{a}^{b}\,(\ln T)_{,b}\ =\ 0,\ \ \ \ \ \ \ \
h_{a}^{b}=u_{a}u^{b}+\delta _{a}^{b}\ ,  
\label{eq:tolma}
\end{equation}

\noindent
leading to

\begin{equation}
\frac{B^{\prime }}{B}\ +\ \frac{T^{\prime }}{T}\ =\ 0\qquad \Rightarrow
\qquad T\ \propto \ B^{-1}  
\label{eq:Ttolman}
\end{equation}

\noindent
The particle number density $n$ trivialy satisfies the conservation law 
$J^{a}\,_{;a}=0$ where $J^{a}=n\,u^{a}$, thus the number of dark particles is 
conserved. Notice that given (\ref{eq:rp}), the equation of state 
(\ref {eq:mb}) and the temperature from the Tolman law (\ref{eq:Ttolman}), 
we have two different expressions for $n$

\begin{equation}
n\ =\ \frac{p}{k_{_{B}}T}\ \propto \ p\,B,  
\label{eq:n1}
\end{equation}

\begin{equation}
n\ =\ \frac{1}{mc^{2}}\,\left[ \rho -\frac{3}{2}\,p\right] ,  
\label{eq:n2}
\end{equation}

The quantity $mc^{2}\,n$ in (\ref{eq:n2}) follows directly from equations 
(\ref{eq:rp})

\begin{equation}
\kappa \,mc^{2}n={\frac{\left[ -8\,vv^{\prime }\,r+\left( {v}^{2}+9\right)
\left( 2\,{v}^{2}+1\right) \right] M+4\,vv^{\prime }\,{r}^{2}-{v}^{2}\left(
7+{v}^{2}\right) r}{{r}^{3}\left( 1+{v}^{2}\right) }},  
\label{eq:rn}
\end{equation}

\noindent
while $n$ in (\ref{eq:n1}) also follows from $p$ in (\ref{eq:rp}) with 
$B\propto \exp [\int {(v^{2}/r)dr}]$. Consistency requires that (\ref{eq:n1})
and (\ref{eq:n2}) yield the same expression for $n$.

\setcounter{equation}{0}
\section{Dark fluid hydrodynamics}

So far we have expressed the field equations and the thermodynamics of the 
fluid source in terms of the tangential velocities $v(r)$ and the effective 
mass-energy $M(r)$. If the dark matter component dominates the dynamics of 
the fluid, we can ignore the contribution from visible matter (barions) and 
assume that the matter source is made exclusively of this dark matter 
component. A useful strategy to follow is then to prescribe, as a functional 
form for $v(r)$, the empiric {\it ansatz} of the radial profile of tangential 
velocities of dark matter halo obtained from the ``universal rotation curve'' 
(URC) that roughly fits observed galactic rotation curves \cite{urc_1}, 
\cite{urc_2}. This empirical form can be used in equations (\ref{eq:const}) 
and (\ref{eq:rp}). The function $M(r)$ follows by solving the constraint 
(\ref{eq:const}), subjected to the appropriate boundary conditions. The 
obtained $M$ together with the prescribed $v$ yield fully determined forms 
for $\rho $, $p$ and the metric coefficient $B(r)$. Next we can verify 
compatibility of $n$ obtained from (\ref{eq:n1}) and (\ref{eq:n2}). Finaly, 
we should be able to estimate the ratio $\beta =mc^{2}/k_{_{B}}T$ which in 
turn, from estimations of masses from particle physics, leads to an
estimation of the temperature of the dark gas in terms of the particles'
mass.

We shall assume for $v^{2}$ the empiric dark halo rotation velocity law given 
by Persic and Salucci \cite{urc_1}, \cite{urc_2}

\begin{equation}
v^{2}\ =\ \frac{v_{0}^{2}\,x^{2}}{a^{2}+x^{2}},\qquad x\ \equiv \ \frac{r}{%
r_{\hbox{\tiny{opt}}}}  
\label{eq:v}
\end{equation}

\noindent
where $r_{\hbox{\tiny{opt}}}$ is the ``optical radius'' containing 83 \% of
the galactic luminosity, whereas the empiric parameters $a$ and $v_{0}$, 
respectively, the ratio of ``halo core radius'' to $r_{\hbox{\tiny{opt}}}$
and the ``terminal'' rotation velocity, depend on the galactic luminosity. 
For spiral galaxies we have:
$v_{0}^{2}=v_{\hbox{\tiny{opt}}}^{2}(1-\beta_*) (1+a^{2})$, where
$v_{\hbox{\tiny{opt}}}=v(r_{\hbox{\tiny{opt}}})$ and the  best fit to
rotation curves is obtained for: $a=1.5\,(L/L_{\ast })^{1/5}$  and
$\beta_* =0.72+0.44\log _{10}(L/L_{\ast })$, where 
$L_{\ast}=10^{10.4}L_{\odot }$. The range of these parameters for spiral 
galaxies is $125\,\hbox{km/sec} \leq v_{0} \leq 250\,\hbox{km/sec}$ and 
$0.6 \leq a \leq 2.3$ \footnote{The URC given by (\ref{eq:v}) fits also 
elliptic and irregular galaxies.}. The field equation (\ref{eq:const}) 
becomes the following linear first order ODE

\begin{eqnarray}
&&\frac{dM}{dx}\ +\ {\frac{\left( 1+2\,{{\it v_{0}}}^{2}\right) \left( 
{{\it v_{0}}}^{2}-3\right) {x}^{4}-{a}^{2}\left( {{\it v_{0}}}^{2}+6\right) 
{x}^{2}-3\,{a}^{4}}{x\left( {b}^{2}+{x}^{2}\right) \left( {a}^{2}+{x}^{2}
+{{\it v_{0}}}^{2}{x}^{2}\right) }}\,M\ +  \nonumber \\
&&+\ {\frac{{{\it v_{0}}}^{2}{x}^{4}\left( 2-{{\it v_{0}}}^{2}\right) \,r_{
\hbox{\tiny{opt}}}}{\left( {a}^{2}+{x}^{2}\right) \left( {a}^{2}+{x}^{2}+{
{\it v_{0}}}^{2}{x}^{2}\right) }}\ =\ 0,  
\label{eq:M1}
\end{eqnarray}

\noindent
the state variables (\ref{eq:rp}) become

\begin{equation}
\frac{1}{2}\kappa \,\rho \ =\ {\frac{\left[ \left( 1+2\,{{\it v_{0}}}
^{2}\right) \left( 3-{{\it v_{0}}}^{2}\right) {x}^{4}+{a}^{2}\left( {{\it 
v_{0}}}^{2}+6\right) {x}^{2}+3\,{a}^{4}\right] M-{{\it v_{0}}}^{2}\left( 2-{
{\it v_{0}}}^{2}\right) {x}^{5}\,r_{\hbox{\tiny{opt}}}}{\left( {a}^{2}+{x}
^{2}\right) \left[ {a}^{2}+(1+{{\it v_{0}}}^{2})\,{x}^{2}\right] {x}^{3}\,r_{
\hbox{\tiny{opt}}}^{3}}},  
\label{eq:rho2}
\end{equation}

\begin{equation}
\frac{1}{2}\kappa \,p\ =\ {\frac{-\left[ {a}^{2}+\left( 1+2\,{{\it v_{0}}}
^{2}\right) {x}^{2}\right] M+{{\it v_{0}}}^{2}{x}^{3}\,r_{\hbox{\tiny{opt}}}
}{\left( {a}^{2}+{x}^{2}\right) {x}^{3}\,r_{\hbox{\tiny{opt}}}^{3}}},
\label{eq:p2}
\end{equation}

\noindent
while the metric coefficient $B$ takes the form

\begin{equation}
B\ =\ \left[ 1+x^{2}\right] ^{v_{0}^{2}/2}\qquad \Rightarrow \qquad T\ =\
T_{c}\,\left[ 1+x^{2}\right] ^{-v_{0}^{2}/2},  
\label{eq:AT}
\end{equation}

\noindent
so that $T$ is the temperature complying with the Tolman law and $T_{c}=T(0)$.
Given a solution $M=M(x)$ of (\ref{eq:M1}), all state variables become 
determined as functions of $x$ and $v_{0}$. The solution of (\ref{eq:M1}) is 
the following quadrature

\begin{equation}
M\ =\ \frac{(v_{0}^{2}-2)(a^{2}+x^{2})^{2-v_{0}^{2}}\,v_{0}^{2}\,x^{3}}{
\left[ a^{2}+(1+v_{0}^{2})x^{2}\right] ^{2/(1+v_{0}^{2})}}\,r_{
\hbox{\tiny{opt}}}\,\int {\frac{\left[ a^{2}+(1+v_{0}^{2})x^{2}\right]
^{(1-v_{0}^{2})/(1+v_{0}^{2})}\,x\,dx}{(a^{2}+x^{2})^{3-v_{0}^{2}}}},
\label{eq:M2}
\end{equation}

\noindent
where we have set an integration constant to zero in order to comply with
the consistency requirement that $v_{0}=0$ implies flat spacetime 
($B=1,\,M=0$). Since the velocities of rotation curves are newtonian, 
$v_{0}\ll c$ (typical values are $v_{0}/c\approx 0.5\times 10^{-3}$), 
instead of evaluating (\ref{eq:M2}) we will expand this quadrature around 
$v_{0}/c$ (in order to keep the notation simple, we write $v_{0}$ instead of 
$v_{0}/c$). This yields

\begin{equation}
M\ =\ \frac{x^{3}\,r_{\hbox{\tiny{opt}}}}{a^{2}+x^{2}}\,v_{0}^{2}\,\left[ 1-
\frac{5x^{2}+2a^{2}}{2(a^{2}+x^{2})}\,v_{0}^{2}+\frac{
12x^{4}+11a^{2}x^{2}+3a^{4}}{2(a^{2}+x^{2})^{2}}\,v_{0}^{4}+{\cal O}
(v_{0}^{6})\right] ,  
\label{eq:M3}
\end{equation}

\noindent
we obtain the expanded forms of $\rho $ and $p$ by inserting (\ref{eq:M2}) 
into (\ref{eq:rho2}) and (\ref{eq:p2}) and then expanding around $v_{0}$,
leading to

\begin{equation}
\kappa \,\rho \,r_{\hbox{\tiny{opt}}}^{2}\ =\ \frac{2(3a^{2}+x^{2})}{
(a^{2}+x^{2})^{2}}\,v_{0}^{2}-\frac{5x^{4}+23a^{2}x^{2}+6a^{4}}{
(a^{2}+x^{2})^{3}}\,v_{0}^{4}+{\cal O}(v_{0}^{6}),  
\label{eq:rho3}
\end{equation}

\begin{equation}
\kappa \,p\,r_{\hbox{\tiny{opt}}}^{2}\ =\ \frac{2a^{2}+x^{2}}{
(a^{2}+x^{2})^{2}}\,v_{0}^{4}-\frac{2x^{4}+7a^{2}x^{2}+3a^{4}}{
(a^{2}+x^{2})^{3}}\,v_{0}^{6}+{\cal O}(v_{0}^{8}),  
\label{eq:p3}
\end{equation}

\noindent
while the expanded form for $T$ follows from (\ref{eq:AT})

\begin{equation}
T\ =\ T_{c}\,\left[ 1-\frac{1}{2}\,\ln \left( 1+x^{2}\right) \,v_{0}^{2}+
\frac{1}{8}\,\ln ^{2}\left( 1+x^{2}\right) \,v_{0}^{4}-{\cal O}(v_{0}^{6})
\right] \ .
\end{equation}

In order to compare $n$ obtained from (\ref{eq:n1}) and (\ref{eq:n2}), we 
substitute (\ref{eq:v}) and (\ref{eq:M2}) into (\ref{eq:rn}) and expand 
around $v_{0}$, leading to

\begin{equation}
n\,r_{\hbox{\tiny{opt}}}^{2}\ ={\frac{{1}}{{\kappa \,m\,c^{2}}}}\,\left[
\frac{2(3a^{2}+x^{2})}{(a^{2}+x^{2})^{2}}\,v_{0}^{2}-\frac{
18x^{4}+55a^{2}x^{2}+13a^{4}}{2(a^{2}+x^{2})^{3}}\,v_{0}^{4}+{\cal O}
(v_{0}^{6})\right] \  ,  
\label{eq:n11}
\end{equation}

\noindent
while $n$ in (\ref{eq:n1}) follows by substituting (\ref{eq:M2}) into 
(\ref{eq:p3}), using $T$ from (\ref{eq:AT}) and then expending around $v_{0}$. 
This yields

\begin{eqnarray}
&&n\,r_{\hbox{\tiny{opt}}}^{2}\ ={\frac{{1}}{{\kappa \,k_{_{B}}T_{c}}}}\left[
\frac{2a^{2}+x^{2}}{(a^{2}+x^{2})^{2}}\,v_{0}^{4}+\right.   \nonumber \\
&&\left. +\frac{(2a^{2}+x^{2})(a^{2}+x^{2})\ln \left( 1+x^{2}\right)
-2(a^{2}+2x^{2})(3a^{2}+x^{2})}{2(a^{2}+x^{2})^{3}}\,v_{0}^{6}+{\cal O}
(v_{0}^{8})\right] \ .  \nonumber \\
&&  
\label{eq:n22}
\end{eqnarray}

\noindent
Since $v_{0}/c\ll 1$, a reasonable approximation is obtained if the leading 
terms of $n$ from (\ref{eq:n11}) and (\ref{eq:n22}) coincide. By looking at 
these equations, it is evident that this consistency requirement implies

\begin{equation}
\frac{1}{2}\,mv_{0}^{2}\ =\ \frac{3}{2}\,k_{_{B}}T_{c}\ ,
\label{eq:consist}
\end{equation}

\noindent
where $v_{0}$ denotes a velocity ($\hbox{km}/\hbox{sec}$) and not the 
adimensional ratio $v_{0}/c$. Equation (\ref{eq:consist}) is analogous to
the condition that defines the so-called ``virialized temperature'' in
the context of cooling of a baryon gas, though in the present approach
such a temperature corresponds to the dark matter gas (see section 17.3 of
\cite{tbook_2}). Since higher order terms in $v_{0}/c$ have a minor
contribution, the two forms of $n$ are approximately equal. This is 
shown in Figure 1 displaying the adimensional quantity 
$\kappa \,mc^{2}n\,r_{\hbox{\tiny{opt}}}^{2}$ from (\ref{eq:n11}) and 
(\ref{eq:n22}) as functions of $x$ for typical values 
$v_{0} = 200\, \hbox{km/sec},\,a=1$ and eliminating $T_{c}$ with
(\ref{eq:consist}).  Equation (\ref{eq:M2}) 
shows how ``flattened'' rotation curves, as obtained from the empiric form 
(\ref{eq:v}), lead to $M\propto r^{3}$ for $r\approx 0$ and $M\propto r$ for 
large $r$. Equations (\ref{eq:rho2}) to (\ref{eq:consist}) represent a 
relativistic generalization of the ``isothermal sphere'' that follows as the 
newtonian limit of an ideal Maxwell-Boltzmann characterized by 
$\rho \approx mc^{2}n$,\thinspace $p\ll \rho $ and $T\approx T_{c}$. In fact, 
using newtonian hydrodynamics we would have obtained only the leading terms 
of equations (\ref{eq:rho2}) to (\ref{eq:consist}). It is still interesting 
to find out that the isothermal sphere can be obtained from General Relativity
in the limit $v_{0}/c\ll 1$ by demanding that rotation curves have a form 
like (\ref{eq:v}). The total mass of the galactic halo, usualy given as $M$ 
evaluated at the radius $r=r_{200}$ (the radius at which $\rho $ is 200 times 
the mean cosmic density). Assuming this density to be 
$\approx 10^{-29}\,\hbox{gm}/\hbox{cm}^{3}$ together with typical values 
$v_{0}=200\, \hbox{km}/\hbox{sec}$ and $a=1$ yields $r_{200}\approx 150$ kpc. 
Evaluating $M$ at this values yields about $10^{17}\,M_{\odot }$, while $M$ 
evaluated at a typical $r_{opt} = 15$ kpc leads to about 
$10^{12}\,M_{\odot }$, an order of magnitude larger than the galactic mass 
due to visible matter. These values are consistent with Refs.
\cite{urc_1}  and \cite{urc_2}. 

\setcounter{equation}{0}
\section{Discussion.}

So far we have found a reasonable approximation for galactic dark matter to 
be described by a self gravitating Maxwell-Boltzmann gas, under the 
assumption of the empiric rotation velocity law (\ref{eq:v}). The following 
consistency relations emerge from equations (\ref{eq:n11}), (\ref{eq:n22}) 
and (\ref{eq:consist})

\begin{equation}
n_c \ \approx \ \frac{3\,v_0^2}{4\pi G\,m\,a^2\,r_{\hbox{\tiny{opt}}}^2}
,\qquad T_c \ \approx \ \frac{m\,v_0^2}{3\,k_{_B}}  
\label{eq:nTc}
\end{equation}

\noindent
hence, bearing in mind that $n\leq n_c$ and $T\approx T_c$, the condition 
(\ref{eq:landau}) for the validity of the MB distribution can be written as

\begin{equation}
\frac{n\,\hbar^3}{(m\,k_{_B}\,T)^{3/2}} \ \leq \ \frac{n_c\,\hbar^3}
{(m\,k_{_B}\,T_c)^{3/2}} \ \ll \ 1,  
\label{landau2}
\end{equation}

\noindent
Inserting (\ref{eq:nTc}) into (\ref{landau2}) yields the condition

\begin{equation}
m \ \gg \ \left[\frac{3^{5/2}\,\hbar^3}{4\pi G\,a^2\,r_{\hbox{\tiny{opt}}
}^2\,v_0}\right]^{1/4},  
\label{aplicab}
\end{equation}

\noindent
a criteria of aplicability of the MB distribution (non-degeneracy) that is
entirely given in  terms of $m$, the fundamental constants $G,\,\hbar $
and the empiric  parameters $v_0$ and $b \equiv a\,r_{\hbox{\tiny{opt}}}$
(the ``terminal''  rotation  velocity and the ``core radius''). For dark
matter dominated galaxies (spiral  and low surface brightness (LSB))
\cite{urc_1} these parameters have a small  variation range:
$r_{\hbox{\tiny{opt}}} \approx 15$\,\hbox{kpc} , 
$1$ kpc \, $\leq b \leq$ \, 5 kpc and $125\, \hbox{km/sec} \leq v_0 \leq 250\, 
\hbox{km/sec}$, 
the constraint (\ref{aplicab}) does provide a tight estimate of the minimal 
value for the mass of the particles under the assumption that these particles 
form a self gravitating ideal dark gas complying with MB statistics. As shown 
in Figure 2, this minimal value lies between $30$ and $60$ eV, thus implying 
that appropriate particle candidates must have a much larger mass than this 
range of values. This minimal bound excludes, for instance, light mass
particles such  as the electron neutrino ($m_{\nu_e} < 2.2$ eV) or the
axion  ($m_A \approx 10^{-5}$ eV). The currently accepted estimations of
cosmological bounds on the sum of masses for the three active neutrino
species is about 
$24$ eV \cite{PDG}, a value that would apparently rule out all neutrino 
flavours. However, recent estimations of these cosmological bounds have 
raised this sum to about $1$ keV \cite{sn2}, hence more massive neutrinos 
could also be accomodated as dark matter particle candidates. Estimates of 
masses of various particle candidates are displayed in Table~1.

Since $T\approx T_{c}$, the consistency condition (\ref{eq:consist})
provides  the following constraint on the temperature and particles mass
of the dark gas

\begin{equation}
\frac{m}{T_{c}}\ =\ \frac{3\,k_{_{B}}}{v_{0}^{2}}\ \approx 0.4\times 10^{3}\,
\frac{\hbox{eV}}{\hbox{K}} \ ,
\label{eq:consist2}
\end{equation}

\noindent
where we have taken $v_{0}=200\,\hbox{km/sec}$ \footnote{The variation of
$v_0$ in the observed ranges for spiral galaxies does not alter significanly
the numerical value in the rhs of Eq. (\ref{eq:consist2}).}. Considering in 
(\ref{eq:consist2}) the minimal mass range that follows from (\ref{aplicab}), 
we would obtain gas temperatures consistent with the assumed typical 
temperatures of relic gases: $T_{c}\approx 2\,\,\hbox{to}\,\,4\,\hbox{K}$. 
However, as long as we do not have more information on the interaction and
physical properties of  various particle candidates, we cannot
rule out a given large mass value on the grounds that the corresponding gas
temperature could be too high. However, if we assume that the ideal  dark
gas is made of electrons or barions, so that $m=m_{p}$ or
$m=m_{e}$,  then condition (\ref{aplicab}) for applicability of the MB
distribution is  certainly satisfied, but (\ref{eq:consist2}) implies a
temperature of the  order of $T_{c}\approx 10^{3}$ K for electrons and
$T_{c}\approx 10^{6}$ K  for barions ! Obviuosly, barions or electrons at such
a high temperatures  would certainly not remain unobservably ``dark''. We can
rule them out, but we cannot rule out more massive particles (in the range of
1-100 GeV's) characterized by weak interaction even if the gas temperature is
in the range of $10^8 - 10^9$ K. Figure 3 illustrates, for various particle
candidates, the relation between $T_c$ and $m$ contained in
(\ref{eq:consist2}).  The main  novelty of the present paper is the fact that
it is based on a general  relativistic hydrodynamics, as opposed to numerical
simulations \cite{nbody_1}-\cite{nbody_3}, newtonian or Kinetic Theory
perturbative  approaches (see \cite{scdm_1}-\cite{wdm_6}).

Finaly, the fact that under the assumption of MB distribution, we have 
obtained a minimal mass on the range 
$30 \,\hbox{eV} \leq m \leq 60 \,\hbox{eV}$
that seems to discriminate against thermal relique gases composed by lighter
particles (electron neutrino, etc) coincides with the fact that
these particle candidates tend to  be ruled out because of their inability
to produce sufficient matter  clustering \cite{tbook_1}, \cite{tbook_2}. In
spite of these  arguments, if either of these particles constitute a self
gravitating gases  accounting for a galactic halo it would be inconsistent
to model such a gas  as SCDM in the context of a classical ideal gas
complying with MB  statistics. It would be necessary to examine these cases
as either HDM or  WDM, by using either a relativistic MB distribution (very
light particles  can be relativistic even at low temperatures) and/or a
distribution that  takes into account (depending on the particle)
Fermi-Dirac or Bose-Einstein  statistics. Non-thermal axions are very
light particles ($m \sim 10^{-5}$ eV), however this type of relique source
cannot be treated as a Maxwell-Boltzmann gas, thus the lower mass limit
that we have obtained does not apply. This and other non-thermal sources
\cite{sfe_1, sfe_2} require a wholy different approach. These studies will
be undertaken in future papers.


\begin{table}[tbp]
\begin{center}
\begin{tabular}{|c|c|c|}
\hline
{SCDM/WDM} & {mass in keV} & {References} \\
{Light Candidates } & {} & {} \\ \hline\hline
{Light Gravitino\hfill} & {$\sim 0.5$} & {\ \cite{g1} } \\
{\hfill} & {$\sim 0.75 - 1.5$ } & {\cite{g2}, \cite{g3} } \\ \hline
{\hfill} & {$\sim 2.6 - 5$} & {\cite{ssn1}} \\
{Sterile Neutrino\hfill} & {$< 40$ } & {\cite{ssn2} } \\
{\hfill} & {$1 - 100$ } & {\cite{ssn3} } \\ \hline
{Standard Neutrinos\hfill} & {$\sim 1$} & {\cite{sn1}, \cite{sn2} } \\ \hline
{Dilaton\hfill} & {$\sim 0.5$} & {\cite{d1} } \\ \hline
{Light Axino\hfill} & {$\sim 100$} & {\cite{a1} } \\ \hline
{Majoron\hfill} & {$\sim 1$} & {\cite{m1}, \cite{m2}, \cite{m3} } \\ \hline
{Mirror Neutrinos\hfill} & {$\sim 1$} & {\cite{mn1}, \cite{mn2} } \\ \hline
\hline
{CDM} & {mass in GeV} & {References} \\
{Heavy Candidates } & {} & {} \\ \hline\hline
{Neutralino\hfill} & {$> 32.3$} & {\ \cite{abreu} } \\
{\hfill} & {$> 46$ } & {\cite{ellis} } \\ \hline
{Axino\hfill} & {$\sim 10$} & {\cite{covi}, \cite{leszek}} \\ \hline
{Gravitino\hfill} & {${\buildrel <\over \sim }\ 100$} 
& {\cite{kawasaki} } \\ \hline
\end{tabular}
\end{center}
\caption{{\em Particle candidates for a MB Dark Matter gas.}}
\end{table}

\section{Acknowledgments}
We thank Professor Rabindra N. Mohapatra for calling our attention to the 
important papers of Ref. \cite{scdm_4} and \cite{scdm_5} and N. Fornengo
for useful discussions. R. A. S. is partly supported by the {\bf DGAPA-UNAM}, 
under grant (Project No. {\tt IN122498}), T. M. is partly supported by 
{\bf CoNaCyT} M\'exico, under grant (Project No. {\tt 34407-E}) and 
L. G. C. R. has been supported in part by the {\bf DGAPA-UNAM} under grant 
(Project No. {\tt IN109001}) and in part by the {\bf CoNaCyT} under grant 
(Project No. {\tt I37307-E}).

\newpage

\begin{figure}[htb]
\centerline{
\epsfig{figure=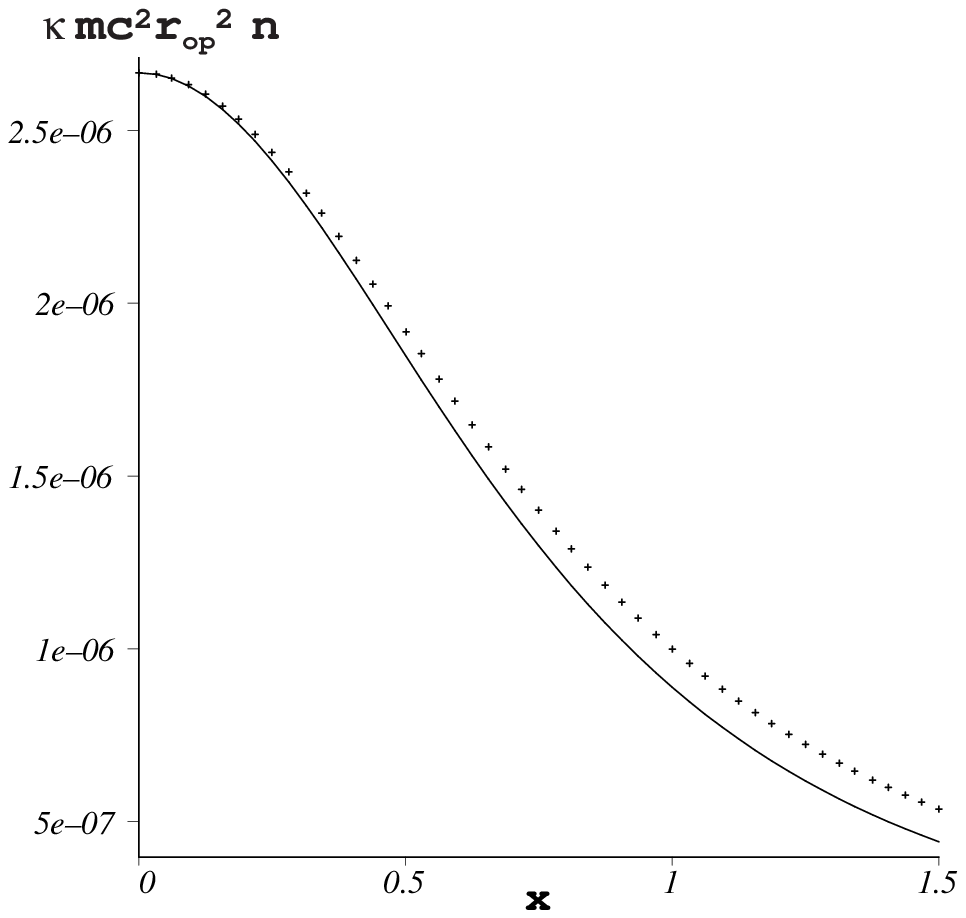,width=4in}}
\caption{$\underline{\hbox{Comparison of $n$ obtained from (\ref{eq:n11}) 
and (\ref{eq:n22})}}$. This plot displays the adimensional quantity 
$\kappa m c^2 r_{op}^2\,n$, as a function of $x$, obtained from truncating 
the right hand sides of (\ref{eq:n11}) (solid curve) and (\ref{eq:n22}) 
(dotted curve with crosses) up to fourth order in $v_0/c$ and  assuming 
the consistency condition (\ref{eq:consist}). Since we are using an empiric 
law for observed galactic rotation curves (the URC given by (\ref{eq:v})), 
the fact that these two expressions for $n$ are so close to each other 
provides an empirical justification for the compatibility between MB 
distribution and these rotation curves.}
\label{fig1}
\end{figure}

\newpage

\begin{figure}[htb]
\centerline{
\epsfig{figure=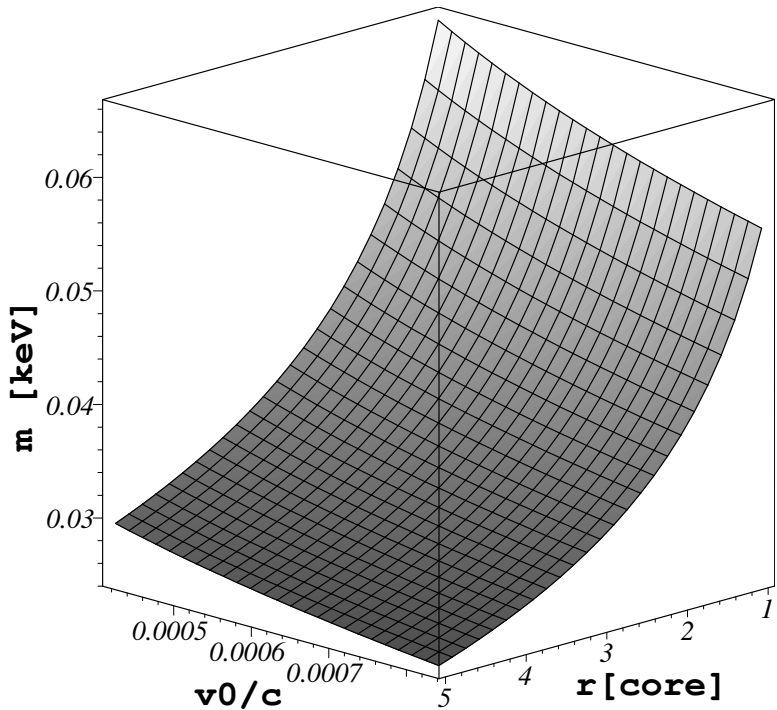,width=4in}}
\caption{$\underline{\hbox{Minimal mass for which the Maxwell-Boltzmann
distribution is applicable}}$.  This graph displays $m$ (in keV's) as a 
function of $v_0/c$ and $b=a\,x$, respectively, the terminal velocity and 
`halo core radius' associated with the URC given in (\ref{eq:v}). Assuming 
typical ranges for spiral galaxies: 
$125\,\hbox{km/sec} \leq v_0 \leq 250\,\hbox{km/sec}$ and 
$1\,\hbox{kpc} \leq b \leq 5\,\hbox{kpc}$, we obtain masses in the range of 
$30 \,\hbox{eV} \leq m \leq 60 \,\hbox{eV}$ that follow from the right hand 
side of the relation (\ref{aplicab}), providing the criterion for 
applicability of the Maxwell-Bolzmann distribution. Dark matter particle 
candidates complying with an MB distribution must have much larger mass 
than the plotted values\,  $30 \,\hbox{eV} \leq m \leq 60 \,\hbox{eV}$.}
\label{fig2}
\end{figure}

\newpage

\begin{figure}[htb]
\centerline{
\epsfig{figure=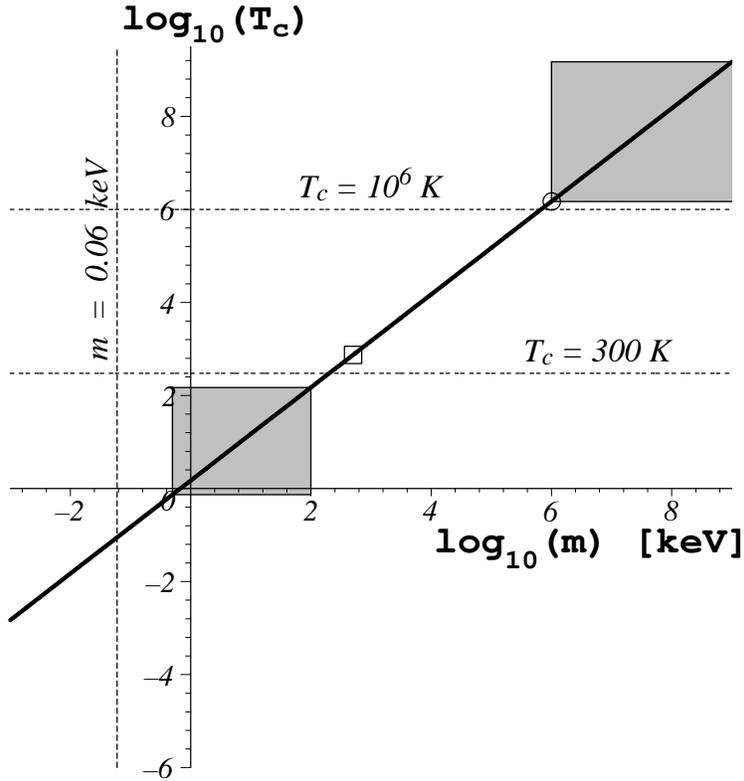,width=4in}}
\caption{$\underline{\hbox{Relation between particle mass and central
temperature}}$. This graph displays the relation between $\log_{10}(T_c)$
(in K) and $\log_{10}(m)$ (in keV's) that follows from equation 
(\ref{eq:consist2}) for a terminal velocity $v_0=200\,\hbox{km/sec} $. 
Almost identical plots are obtained for other velocities in the observed 
range $125 \,\hbox{km/sec} \leq v_0 \leq 250 \,\hbox{km/sec}$. The circle 
and box symbols respectively denote the proton and electron mass yielding 
central temperatures of the order $T_c\approx 10^6,\,10^3 $ K. The central 
temperature for light particles in the range 
$0.5 \,\hbox{keV} \leq m \leq 100 \,\hbox{keV}$ is less than $300$ K 
(rectangle in the left), while for massive sypersymmetric particles in the 
range $1 \,\hbox{GeV} \leq m \leq 100 \,\hbox{GeV}$, we have $T_c$ as large 
as $10^9$ K (rectangle on the right). However, such high temperatures cannot 
rule out these weakly interactive particles as components of the dark matter 
MB gas. }
\label{fig3}
\end{figure}

\end{document}